\documentclass[osajnl,preprint]{revtex4-1}

\usepackage{amsmath}
\usepackage{amsfonts}
\usepackage{graphicx}
\usepackage{epstopdf}
\usepackage{hyperref}

\begin{document}

\title{Non-diffracting chirped Bessel waves in optical antiguides}

\author{Ioannis Chremmos}
\email{ioannis.chremmos@mpl.mpg.de}
\affiliation{Max Planck Institute for the Science of Light, D-91058 Erlangen, Germany}

\author{Melpomeni Giamalaki}
\affiliation{School of Electrical and Computer Engineering, National Technical University, GR 157-73, Athens, Greece}


\begin{abstract}
Chirped Bessel waves are introduced as stable (non-diffracting) solutions of the paraxial wave equation in optical antiguides with a power-law radial variation in their index of refraction. Through numerical simulations, we investigate the propagation of apodized (finite-energy) versions of such waves, with or without vorticity, in antiguides with practical parameters. The new waves exhibit a remarkable resistance against the defocusing effect of the unstable index potentials, outperforming standard Gaussians with the same full width at half maximum. The chirped profile persists even under conditions of eccentric launching or antiguide bending and is also capable of self-healing like standard diffraction-free beams in free space.
\end{abstract}

\maketitle

\section{Introduction}
Structured light waves have become particularly interesting in recent years due to their remarkable properties and extended range of applications.\cite{AndrewsBook} A big part of the effort in the field has been targeted at the so-called \textit{non-diffracting waves}, namely waves that resist diffraction (in their realistic finite-energy versions) and thus maintain their transverse intensity profile over large distances.\cite{Figueroa2013} Bessel beams \cite{Durnin1987} are perhaps the most familiar example with diverse applications in optical manipulation, atom and nonlinear optics. \cite{Mcgloin_2005} Other examples are Mathieu \cite{Gutierrez-Vega2000} and parabolic beams \cite{Bandres2004} which follow as separable solutions of the wave equation in elliptic or parabolic coordinates, respectively. Self-imaging beams that reproduce their profile periodically with distance and beams that rotate their profile around the axis of propagation also qualify as non-diffracting or, more generally, as propagation-invariant waves.\cite{Piestun_1998} More recently, the family of non-diffracting beams expanded with the introduction of accelerating Airy beams \cite{Siviloglou2007} and, subsequently, with the complete set of paraxial accelerating beams.\cite{Bandres2009} These waves have a non-diffracting profile that shifts laterally according to a quadratic law with the propagation distance, thus writing a parabola in free space. The demonstrated and envisaged applications of accelereting beams have also been numerous.\cite{Hu2012Springer} Hybrids between the accelerating and non-accelerating families are also possible, as for example the accelerating Bessel-like beams in the paraxial \cite{Chremmos_2013_Bessel} or nonparaxial \cite{Chremmos_PRA2014} regime.

Non-diffracting waves owe their properties to their special structure of rays which is such that the resulting transverse interference pattern does not change with the propagated distance. A Bessel beam, for example, is produced by ray cones with a fixed opening angle and expanding circular bases on the input aperture.\cite{Durnin1987} With appropriate deformation of the cones, their apexes can be made to trace a curved path which results in an accelerating Bessel-like beam.\cite{Chremmos_2013_Bessel} Airy beams, on the other hand, have a very different structure since their rays create parabolic optical caustics. What is however common among non-diffracting waves is that their profile at increasing propagation distances is produced by rays emitted from points on the input aperture that are increasingly far from their ``center of mass'' (The center of mass of a beam with amplitude $u$ propagating along the $z$ axis lies at the transverse position $\left( {\int {{\bf{r}}|u|^2 dxdy} } \right)/\left( {\int {|u|^2 dxdy} } \right)$). The on-axis far field of a Bessel beam, for example, is produced by rays emitted from the expanding circles on the input plane mentioned before. This property of non-diffracting beams ensures that, if some of the inner rays are scattered or blocked, the outer rays will still produce the beam's profile at longer distances, which is a remarkable property of non-diffracting beams called \textit{self-healing}.\cite{Garces-Chavez2002, Broky2008}  A wave-optics description of this property is also possible. \cite{Aiello2014} This should be contrasted with Gaussian beams where the on-axis field is produced by the single on-axis ray (except on the focal point). If this ray is blocked then the field along the entire $z$ axis is distorted. Along the same lines, certain advantages have been highlighted in the propagation of non-diffracting beams in turbulent media. \cite{Chu2011_AiryTurbulence, Gu2010_AiryScintillation, Eyyuboglu2013_BesselScintillation}

Perusing the literature one sees that non-diffracting waves have been so far a concept limited in free space or homogeneous media, with medium inhomogeneities introduced only locally or randomly when self-healing after scattering or robustness in turbulence are to be studied, respectively. Going beyond these ideas, one may wish to explore the possibility of waves that propagate in a non-diffracting fashion in non-scattering inhomogeneous media that are inherently more diffractive than free space. That would be for example the case of \textit{optical antiguides} (also called unstable optical ducts \cite{Siegman1986}), namely cylindrically symmetric inhomogeneous media whose index of refraction increases with the distance from the axis thus making rays bend outwards and waves diffract quickly.

That non-diffracting waves of this kind should exist is already suggested by the Airy wavepacket paradigm. The Airy wavefunction is the unique non-spreading solution (excluding the unbounded function Bi) of the one-dimensional (1D) Schr{\"o}dinger equation in a field of constant force.\cite{Berry1979} By the analogy between Schr{\"o}dinger equation and Helmholtz equation for paraxial optical waves, it is deduced that a 1D optical Airy beam propagates without diffracting (and without accelerating!) in a medium whose index of refraction varies linearly in the transverse direction, e.g. $n(x) = n_0 + ax$. This is obvious from the fact that the Airy function solves the differential equation $u''(x) = x u(x)$. Again, the non-spreading nature (or stability) of this wave can be explained by its structure of parabolic rays enveloping a straight caustic.\cite{Berry1979} Now imagine a cylindrical antiguide whose index increases linearly with the distance $\rho$ from the axis. If we consider circularly symmetric waves, their corresponding rays are meridional and the underlying ray equation is identical to the 1D case with the cartesian coordinate $x$ replaced by the radial coordinate $\rho$. Hence, a circularly symmetric beam must exist (a 2D wave) that propagates without diffracting along the unstable linear duct and has in the $\rho - z$ plane the same ray structure with the 1D Airy wavepacket in a linear potential. 

Beyond these simple arguments, in this paper we report a family of waves that propagate without diffracting in optical antiguides with a power-law variation of their refractive index. These waves can have zero or nonzero angular momentum while their radial amplitude satisfies a B{\^o}cher-type differential equation. At sufficiently large radial positions the solution behaves asymptotically like a Bessel function with an argument of the order $\mathcal{O} \left( \rho ^ {(\nu + 2)/2)} \right)$, where $\nu$ is the power with which the refractive index varies. For a certain value of the propagation constant that is equal to the wave number on the antiguide's axis (core wave number), the wave amplitude is exactly proportional to a Bessel function with an argument that is proportional to $\rho ^ {(\nu + 2)/2}$. In any case, this nonlinear dependence of the argument on the radius causes the wave amplitude to oscillate with an increasing spatial frequency similar to a chirped pulse in time. These properties prompt us to name the new waves \textit{chirped Bessel} waves. Similar to standard Bessel beams in homogeneous space, chirped Bessel waves in antiguides carry ideally infinite power which means that their realistic versions have to be subject to some kind of apodization of aperturing.

Closing this introduction, we would like to mention that the question of trapping vortex waves in an optical antiguide was addressed recently. \cite{Marrucci2013} In this work simple vortex modes of the Laguerre-Gauss type were launched in curved antiguides with a Gaussian index profile hoping to observe some trapping effect and guidance of the vortex along the antiguide. The conclusion was negative as the test waves had no special structure (for example chirp) in the radial direction. The chirped Bessel waves introduced here can indeed propagate in a stable (diffraction-resisting) fashion along a straight optical antiguide. Moreover, our simulations show that these waves remain robust when launched in a curved antiguide being however attracted along a curved trajectory toward the direction of increasing index.

\section{Theory}
\subsection{Wave analysis}
\label{Wave analysis}
To keep things simple, we consider paraxial propagation of waves in an optical antiguide with a refractive index that increases radially as
\begin{equation}
n(\rho) = n_0  \left[ 1 + 2 \Delta \left( {\frac{\rho}{a}} \right)^{\nu}  \right] ^{1/2}
\label{refractive_index}
\end{equation}
where $a$ is a characteristic radius and $\Delta > 0$, $\nu \geq 1$ are real parameters. Notice that in Eq. \eqref{refractive_index} we have used a notation that is familiar within the field of graded-index optical fibers.\cite{SeniorFibers} To stay in the paraxial regime and ignore polarization effects, the increase in the refractive index must act as a perturbation to the core index $n_0$. In analogy to the terminology of weakly guiding waveguides and fibers, an antiguide of this form can be thought to be \textit{weakly antiguiding}. With $\Delta \ll 1$, this will be true within distances of few $a$ from the axis. We will return to this point later in this section.

In the paraxial regime, any component of the electromagnetic field can be expressed as $u(\rho, \varphi, z) e^{i k_0 z}$, where $u$ is an envelope function that varies slowly along $z$ and $k_0 = 2 \pi n_0 / \lambda$ is the core wave number at a vacuum wavelength $\lambda$. The evolution equation for the envelope is the paraxial wave (or paraxial Helmholtz) equation which reads
\begin{equation}
2 i k_0 u_z  + u_{\rho \rho }  + \rho ^{ - 1} u_\rho   + \rho ^{ - 2} u_{\varphi \varphi }  - V( \rho ) u = 0
\label{paraxial_equation}
\end{equation}
where $V(\rho) = - 2 k_0 ^2 \Delta (\rho / a) ^{\nu}$. The above equation is often viewed from a quantum-mechanical perspective namely as Schr{\"o}dinger equation for wave function $u(\rho,\varphi, z)$ with $z$ playing the role of time. In this context an antiguide is equivalent to an unstable potential $V(\rho)$ which justifies the minus sign that we used in its definition. We now assume a non-diffracting solution of Eq. \eqref{paraxial_equation} of the form
\begin{equation}
u = U (\rho) \exp \left( i m \varphi + i k_0 b z \right)
\label{envelope}
\end{equation}
where integer $m$ is the order of vorticity (or topological charge) and $b$ is a dimensionless real number with $|b| \ll 1$ so that the assumption of a slowly varying along $z$ envelope stays true. Substituting into Eq. \eqref{paraxial_equation} we get
\begin{equation}
\rho ^2 U''(\rho)  + \rho U'(\rho)  + \left( \frac{2 k_0^2 \Delta}{a^\nu}\rho^{\nu + 2} -2 b k_0 ^2 \rho ^2 - m^2 \right) U(\rho) = 0
\label{bocher_equation}
\end{equation}
with the primes indicating ordinary differentiation with respect to $\rho$. The above falls under the general family of B{\^o}cher's equations \cite{ZwillingerCRC} and has no tabulated solution for a general value of $\nu$. The solution has to be obtained numerically or in the form of a series, a method that is frequently applied to graded-index optical fibers.\cite{SnyderLove} In the particular case $\nu = 2$ and $b<0$, Eq. \eqref{bocher_equation} is called the \textit{Bessel wave} equation and has been studied to some extent in the literature. In analogy to the terminology of Bessel functions, its non-singular solution is termed Bessel wave function of the first kind $\mathcal{J}_m \left( k_0 \sqrt{2 \Delta}/ a, k_0 \sqrt{-2b}, \rho \right)$. \cite{WillatzenVoon2011} 

Let us first note two cases where Eq. \eqref{bocher_equation} admits a solution in terms of standard Bessel functions. The first is the limit $\Delta = 0$ whereby the medium is homogeneous with an index $n = n_0$. Then Eq. \eqref{bocher_equation} reduces to Bessel equation with the nonsingular solution being the Bessel function of the first kind \cite{AbramowitzStegun}
\begin{equation}
U (\rho) = J_m \left(k_0 \sqrt{- 2b} \: \rho \right)
\label{Delta_eq_0}
\end{equation}
We thus obtain a standard Bessel beam of order $m$. Obviously we must have $b<0$, which means that the propagation constant $k_z = k_0 (1 + b)$ is lower than the medium wave number $k_0$ (recall the $e^{i k_0 z}$ factor muliplying the envelope function $u$ of Eq. \eqref{envelope}).

The second case is that of a wave with $b = 0$ namely with a propagation constant $k_z = k_0$ equal to the core wave number. We then obtain a transformed version of Bessel equation with nonsingular solution \cite{Bowman_Bessel}
\begin{equation}
U (\rho) = J_{\frac{2 m}{\nu+2}} \left( k_0 \sqrt{\frac{8 \Delta}{a ^{\nu}}} \frac{\rho ^{(\nu + 2)/2}}{\nu + 2} \right)
\label{bessel_chirped}
\end{equation}
This is a chirped Bessel beam with vorticity $m$ that propagates without spreading (or diffracting) along the antiguide. Note that, for real $\nu$ and integer $m$, the order of the Bessel function $\frac{2 m}{\nu + 2}$ is generally a real number which reduces to $m$ only when $\nu = 0$ namely in a homogenous medium. Why such a solution is chirped we can easily understand from the large-argument asymptotics of Bessel functions
\begin{equation}
J_n(x) \sim \sqrt{\frac{2}{\pi x}} \cos \left( x - \frac{\pi}{4} - \frac{n \pi}{2} \right)
\label{bessel_asymptotic}
\end{equation}
which holds generally for real order $n$ (and complex argument with $|arg(x)|< \pi$).

Having a closed-form solution like \eqref{bessel_chirped} at hand, few interesting remarks are in order. First, note that the argument of the Bessel function in Eq. \eqref{bessel_chirped} can be intuitively explained by the concept of the local radial wave number $k(\rho)$. This is a measure of the spatial frequency of the radial oscillations and, due to the asymptotic form \eqref{bessel_asymptotic}, it is defined as the derivative of the argument in Eq. \eqref{bessel_chirped} with respect to $\rho$. The result is
\begin{equation}
k\left( \rho  \right) = k_0 \sqrt {\frac{{2\Delta }}{{a^\nu  }}} \: \rho ^{\nu /2} = \frac{{2\pi }}{\lambda }\sqrt {n^2 \left( \rho  \right) - n_0^2 } 
\label{local_wavenumber}
\end{equation}
where we have also used Eq. \eqref{refractive_index}. Thus $k(\rho)$ and $k_z$ are connected through the dispersion relationship of a plane wave in a homogeneous medium with index equal to the local value $n(\rho)$. Second, note that in a homogeneous medium with index $n = n_0$, $k_0$ is a cut-off propagation constant for propagating waves in the sense that $k_z < k_0$. As $k_z \to k_0 ^-$, the transverse wave number tends to zero and the corresponding Bessel beam oscillates so slowly that it effectively assumes a constant amplitude which is nonzero only for the $J_0$ case. However, when the index is perturbed as in Eq. \eqref{refractive_index}, waves with $k_z > k_0$ are enabled and, no matter how small $\Delta$ is, chirped Bessel waves can oscillate many times as long as the characteristic radius is large enough compared to the wavelength in the medium $(a \gg \lambda / n_0).$ This is easily deduced from Eq. \eqref{bessel_chirped} by setting $\rho = a$. The argument becomes proportional to $k_0 a \sqrt{\Delta}$ thus allowing a large ratio $a n_0 / \lambda$ to compensate for a small $\Delta$.

It is also interesting to see how the solution of Eq. \eqref{bocher_equation} behaves near the axis of the antiguide. In the limit $\rho \to 0$, the power term $\rho ^{\nu+2}$ becomes negligibly small compared to the quadratic term and Eq. \eqref{bocher_equation} reduces to the Bessel or modified Bessel equation depening on whether $b<0$ or $b>0$, respectively. Thus in this limit the solution behaves either like a Bessel or modified Bessel function of the first kind with the same argument with Eq. \eqref{Delta_eq_0}. This could also be expected intuitively since, near the axis, the medium has a flat index profile $n \approx n_0$ hence the wave should behave as a standard Bessel beam in a homogeneous medium. In any case the small argument asymptotic of Bessel functions reads \cite{AbramowitzStegun}
\begin{equation}
J_n \left( x \right) \sim \frac{{\left( {x/2} \right)^n }}{{\Gamma \left( {n + 1} \right)}}
\label{bessel_small_argument}
\end{equation}
hence the envelope of the wave behaves near the axis as $u \sim ct. \times \rho ^ m e^{\pm i m \varphi}$. In the other extreme, namely as $\rho \to \infty$, the power term $\rho ^{\nu+2}$ dominates the quadratic term which can be neglected. We thus obtain the modified Bessel equation of the $b=0$ case and, since we are far from $\rho = 0$, the solution behaves asymptotically as a linear superposition of a Bessel and a Neumann function (or Bessel function of the second kind) with the argument of Eq. \eqref{bessel_chirped}.

\begin{figure}[t]
\centering \includegraphics[width=0.8\textwidth]{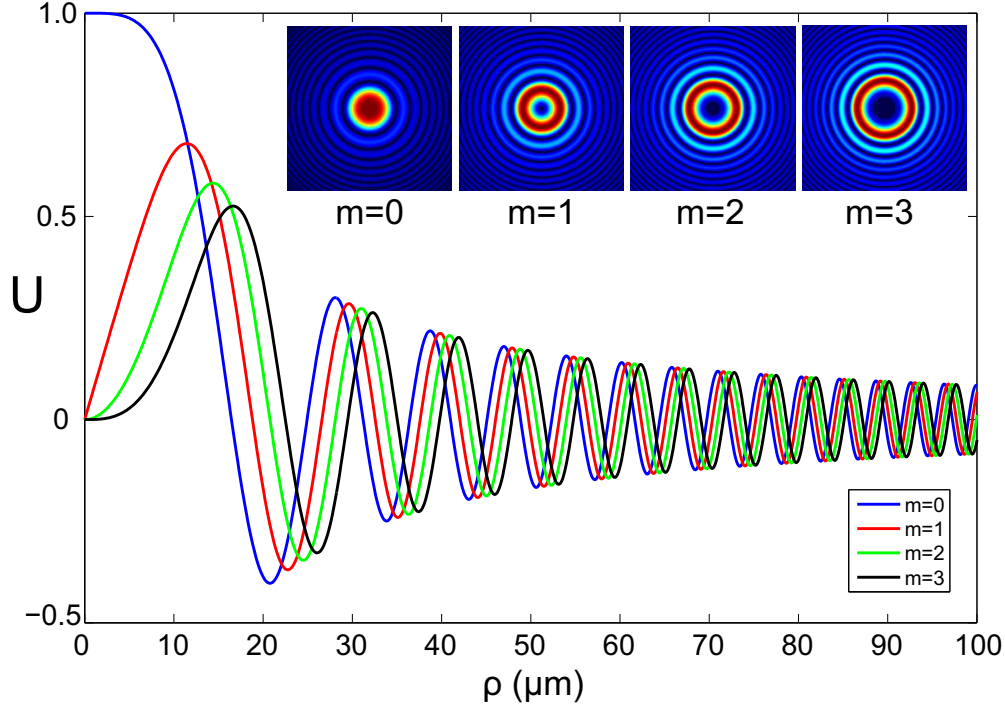}
\caption{(a) Amplitude $U(\rho)$ of the chirped Bessel waves of Eq. \eqref{bessel_chirped} for different orders of vorticity $m = 0, 1, 2, 3$. The insets show the corresponding transverse intensity profiles $|U|^2$. For the antiguide we have chosen the parameters $n_0 = 1.5$, $\Delta = 0.01$, $\nu = 2$, $a = 50 \mu m$ at a vacuum wavelength $\lambda = 1.5  \mu m$.}
\label{Figure_1}
\end{figure}

Figure \ref{Figure_1} shows examples of chirped Bessel waves with $b=0$ and various orders of vorticity. For the antiguide we have used parameters that are common to graded-index optical fibers. One can observe the chirped oscillations of the solution as well as the discussed behaviour near the axis. In the $m=0$ case notice the flattening of the amplitude near the axis which is due to the $J_0 (ct. \times \rho^2)$ dependence.

\subsection{Ray analysis}
\label{Ray analysis}
Before proceeding to numerical simulations, it is instructive to obtain a better physical intuition of the waves under consideration using ray optics. We have already mentioned that the non-diffracting nature of the 1D Airy wavepacket in a linear potential (constant force field) can be explained through its structure of parabolic rays enveloping a straight caustic. A generalization of this picture to a nonlinear power-law potential provides the ray structure of our chirped Bessel waves. For simplicity we limit ourselves to waves with zero angular momentum $(m=0)$. Then the rays are meridional and the paraxial ray equation is written in the $\rho - z$ plane as \cite{SnyderLove} 
\begin{equation}
n\left( \rho  \right)\frac{{d^2 \rho }}{{dz^2 }} = \frac{{dn\left( \rho  \right)}}{{d\rho }}
\label{ray_equation_general}
\end{equation}
where we have used the fact that the index of refraction does not dependent on $z$. As a result, the ray equation is invariant along the $z$ axis or, using the strict mathematical term, autonomous. This implies that, if $\rho(z)$ is a solution to Eq. \eqref{ray_equation_general}, the shifted along the $z$ axis ray $\rho(z - L)$ is also a solution for any $L$. Based on this property, the ray pattern of a non-diffracting wave is obtained as a continuum of shifted copies of a particular solution of the ray equation for $L$ running ideally from $- \infty$ to $\infty$. By revolving this $\rho - z$ pattern around the $z$ axis, the total 3D ray structure is obtained. A transverse cut of this ray structure at any level $z$ yields the same interference pattern which is what one physically observes as a non-diffracting wave.

By substituting Eq. \eqref{refractive_index} into the ray equation \eqref{ray_equation_general} and rearranging we get
\begin{equation}
\frac{{d^2 \rho }}{{dz^2 }} = \frac{{\Delta \nu }}{{a^\nu  }}\rho ^{\nu  - 1}
\label{Emden_Fowler}
\end{equation}
where in the right-hand side we have approximated a factor $\left[ {1 + 2\Delta \left( {\rho /a} \right)^\nu  } \right]^{ - 1}$ by unity. For $\nu = 1$ (a linear potential), the right-hand side is a constant and we obtain parabolic rays of the form
\begin{equation}
\rho _l \left( z \right) = \frac{\Delta}{2 a}z^2 + C_1 z + C_2
\label{nu_eq_1}
\end{equation}
where $C_1,C_2$ are integration constants while the subscript \textit{l} is a reminder of the linear variation of the refractive index. For $\nu = 2$ (inverse harmonic oscillator potential), the ray equation is solved by exponential rays of the general form
\begin{equation}
\rho _q \left( z \right) = C_1 \exp \left( {\frac{{\sqrt {2\Delta } }}{a}z} \right) + C_2 \exp \left( { - \frac{{\sqrt {2\Delta } }}{a}z} \right)
\label{nu_eq_2}
\end{equation}
where the subscript \textit{q} indicates the quadratic variation of the refractive index. For $\nu \neq 2$, Eq. \eqref{Emden_Fowler} is a second-order nonlinear differential equation of the Emden-Fowler type \cite{ZwillingerCRC} and has a known implicit solution. \cite{Polyanin_ODEs} Instead of directly applying the relevant formula, we here find it more useful to reduce Eq. \eqref{Emden_Fowler} to an equivalent first-order differential equation. This is done by multiplying both sides with $d\rho / dz$ and integrating with respect to $z$. The result is
\begin{equation}
\frac{{d\rho }}{{dz}} =  \pm \sqrt {2\Delta \left( {\frac{\rho }{a}} \right)^\nu + C_3}
\label{ray_equation_firstorder_1}
\end{equation}
where $C_3$ is a dimensionless constant of integration. This constant is connected with the propagation constant of our non-diffracting waves. This is better understood if one views Eq. \eqref{ray_equation_firstorder_1} as the conservation law of the longitudinal momentum in our system. Indeed, in a medium that is invariant along $z$, the $k_z$ component of the wavevector of a ray is conserved exactly as the tangential wave vector after refraction through a planar interface (Snell's law). It follows that each ray is characterized by a $k_z$ that stays invariant along its trajectory and is given by the projection of the local wave vector along the $z$ axis, namely 
\begin{equation}
k_z  = \frac{{2\pi }}{\lambda }n\left( \rho  \right)\cos \theta _z  = \frac{{2\pi }}{\lambda }n\left( \rho  \right)\left[ {1 + \left( {\frac{{d\rho }}{{dz}}} \right)^2 } \right]^{ - 1/2} 
\label{ray_kz}
\end{equation}
where $\theta _z= \arctan{(d \rho / dz)}$ is the angle between the ray and the $z$ axis. According to our definition \eqref{envelope} and the previous discussion, the rays that compose our non-diffracting waves have the invariant $k_z = k_0 (1 + b)$. Substituting into Eq. \eqref{ray_kz} and making a paraxial approximation $(|d \rho / dz| \ll 1)$, we get
\begin{equation}
\frac{{d\rho }}{{dz}} =  \pm \sqrt {2\Delta \left( {\frac{\rho }{a}} \right)^\nu -2b}
\label{ray_equation_firstorder_2}
\end{equation}
which is identical with Eq. \eqref{ray_equation_firstorder_1} if $C_3 = -2b$.

The particular form of the ray equation \eqref{ray_equation_firstorder_1} or \eqref{ray_equation_firstorder_2} allows to distinguish two different regimes or patterns of ray structure. For $b>0$, the slope of a ray with respect to the axis $d\rho / dz$ vanishes at the radius $\rho _c  = a\left( { \frac{b}{\Delta} } \right)^{1/\nu }$. This radius is a turning point of the rays and is identified with a caustic radius of the corresponding non-diffracting wave. Since the rays cannot reach this region, the intensity of the wave is low in $\rho < \rho _c$. For waves with $b<0$, on the other hand, $d\rho / dz$ never vanishes and the rays cross the axis of the antiguide with a slope $\pm \sqrt{-2b}$. After the crossing their convexity changes sign and the total trajectory is symmetric with respect to the crossing point. In the boundary between the two regimes, $b=0$ and the ray equation can be integrated analytically to yield
\begin{equation}
\rho _{b=0} \left( z \right) = a\left[ \frac{\sqrt{2\Delta}}{a} \left( \frac{\nu }{2} - 1 \right) {\left( L \pm  z \right)} \right] ^{\frac{2}{{2 - \nu}}}
\label{C_eq_0}
\end{equation}
where the sign is chosen to yield a positive number inside the brackets. Thus the rays either touch the $z$ axis with zero slope when $\nu < 2$, either they are asymptotic to the axis when $\nu > 2$. In the latter case the rays are also asymptotic to transverse planes $z=\pm L$. Finally, for $b=0$ and $\nu = 2$, Eq. \eqref{ray_equation_firstorder_1} yields exponential rays which are also asymptotic to the $z$ axis. These rays are also obtained from Eq. \eqref{nu_eq_2} for $C_1 = 0$ or $C_2 = 0$. 

\begin{figure}[t]
\centering \includegraphics[width=0.9\textwidth]{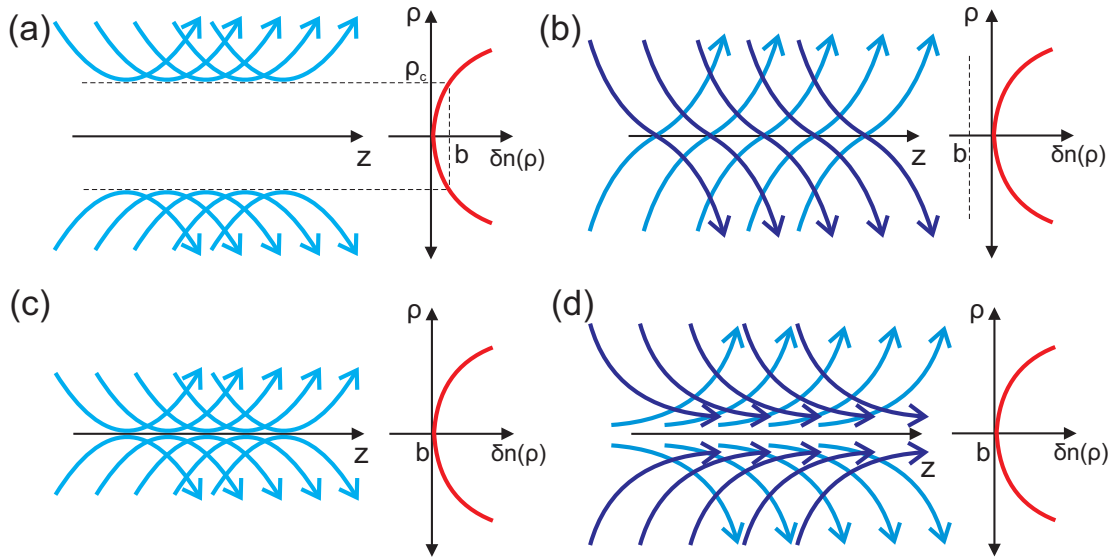}
\caption{Ray patterns of non-diffracting waves in an optical antiguide for (a) $b>0$, (b) $b<0$, (c) $b=0$ and $\nu < 2$, (d) $b=0$ and $\nu \geq 2$. The red curves on the right of the patterns are the relative index difference $\delta n(\rho) = (n(\rho) - n_0)/n_0$ with the value of $b$ indicated. In (b) and (d) two different ray colors are used for visualization purposes.}
\label{Figure_2}
\end{figure}

Figure \ref{Figure_2} illustrates the different cases of rays discussed above. The rays are repeated along the $z$ axis to show the ray structure of our chirped Bessel waves in a meridional plane. The slope of the rays has been exagerrated for visualization purposes. Figure \ref{Figure_2}(a) shows the case $b>0$ where the rays are reflected at the caustic radius $\rho _c$ which is the point at which $b$ equals the relative index difference $\delta n(\rho) \approx \Delta (\rho / a)^{\nu}$. In (b) we have the case $b<0$ where the rays cross the axis and continue with the opposite convexity. In (c) we have the case $b=0$ and $\nu < 2$ in which the rays are reflected at the axis, while (d) shows the case $b=0$ and $\nu \geq 2$ where the rays are asymptotic to the axis. Finally note that these ray patterns suggest that the chirped Bessel waves are also able to self-heal like their standard Bessel counterparts, since the rays contributing to the field near the axis at a large distance $z$ are emitted from far off-axis points on the input plane $z=0$.

The above concern waves without angular momentum. By introducing the azimuthal coordinate in the ray equation, ray patterns of vortex waves can also be derived. These involve skew rather than meridional rays. The vorticity introduces an additional unstable potential with a dependence $m^2 / \rho ^2$ which acts as a centrifugal force pushing the intensity distribution of the waves away from the axis which becomes a field node. Finally note that, in using paraxial ray optics results, one should be aware of their limitations. A rule of thumb is that the paraxial approximation is satisfactory for ray angles smaller than $30^0$ or roughly slopes $|d\rho/dz| < 0.5$. For larger angles, the complete nonparaxial ray equation is required to determine the accurate shapes of the rays. \cite{SnyderLove} For example, rays in a medium with a linear index profile are actually parabolic only within a finite propagation length around their turning point, where the paraxial approximation holds, while they diverge with an exponential law at larger slopes. However, we will not pursue ray optics any further here.

\section{Numerical experiments}
We now proceed to investigate numerically the propagation of chirped Bessel waves in antiguides. As mentioned, their radial profile is obtained by solving Eq. \eqref{bocher_equation}. To this end, we transform this second-order differential equation to a system of two first-order differential equations for functions $U(\rho)$ and $U'(\rho)$ and then we solve this system using a high-order Runge-Kutta scheme.\cite{Griffiths_ODE} To avoid the singularity of this system at $\rho = 0$, the initial conditions are imposed at $\rho = \epsilon > 0$ and their values are obtained from the limiting behaviour of the solution that we determined in section \ref{Wave analysis}. For a wave with $m=0$ for example, we may start with $U(\epsilon)=1$, $U'(\epsilon)=0$.

\begin{figure}[t]
\centering \includegraphics[width=0.9\textwidth]{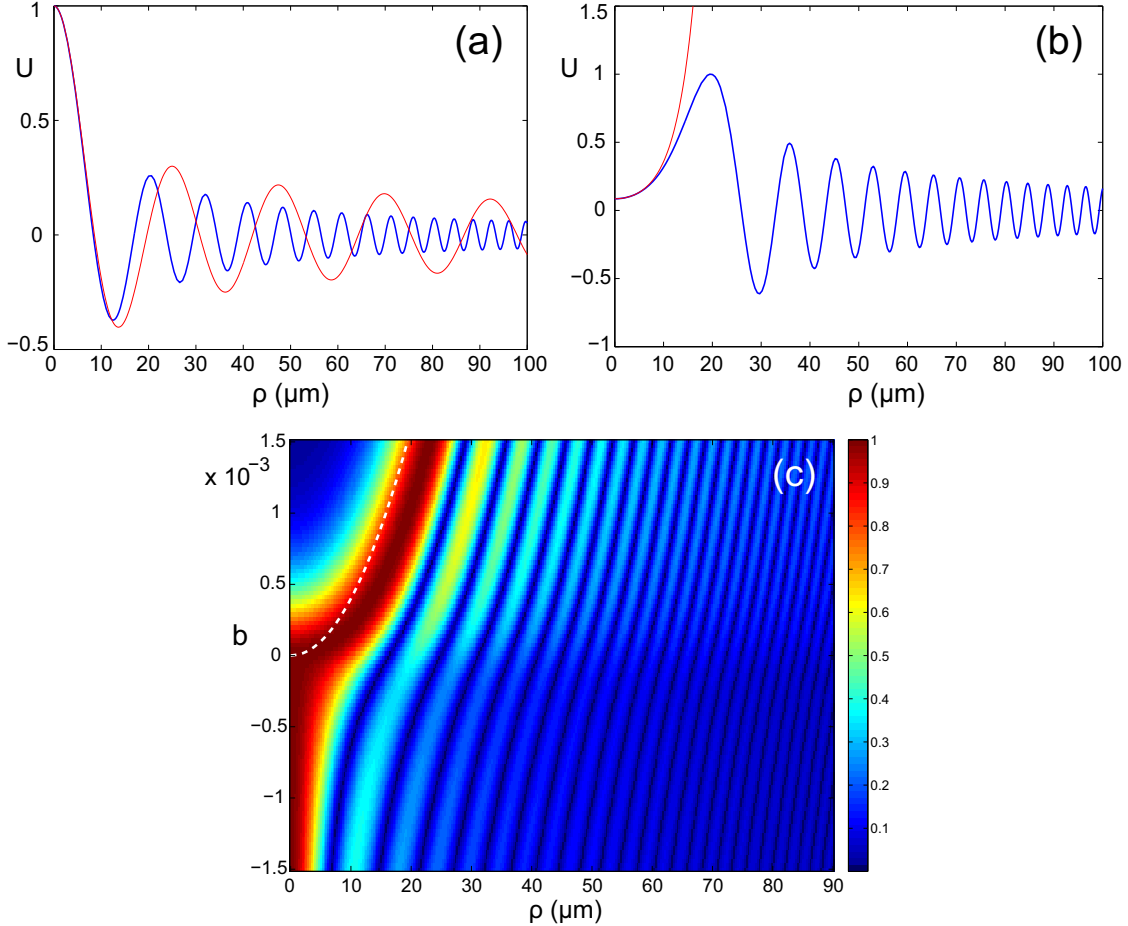}
\caption{Radial profiles (blue curves) of chirped Bessel beams obtained by solving Eq. \eqref{bocher_equation} numerically with (a) $b = -0.001$ $(k_z = 0.999 k_0)$ and (b) $b = 0.001$ $(k_z = 1.001 k_0)$. The thin red curves are the Bessel functions that are asymptotic to the solution as $\rho \to 0$ given by Eq. \eqref{Delta_eq_0}.  (c) A continuous map of the magnitude $|U|$ versus $\rho$ and $b$. For any $b$ the maximum intensity has been scaled to unity. The dashed white curve is the caustic radius $\rho _c = a (b / \Delta)^{1/ \nu}$. The antiguide's parameters and the operating wavelength are the same with Fig. \ref{Figure_1}.}
\label{Figure_3}
\end{figure}

Figures \ref{Figure_3}(a) and (b) show the radial profile of chirped Bessel waves with $m=0$ and $b<0$ $(k_z < k_0)$ or $b>0$ $(k_z > k_0)$, respectively. The parameters of the antiguide and the operating wavelength are the same with Fig. \ref{Figure_1}. For $b<0$ the wave resembles a zero-order Bessel beam with a maximum on axis and oscillations that are getting faster with increasing $\rho$. For $b>0$, the maximum shifts away from the axis which reflects the existence of a caustic radius, namely the turning point of the rays that was discussed in Fig. \ref{Figure_2}(a). In both cases the oscillations are linearly chirped (phase increasing with $\rho ^2$) as expected from Eq. \eqref{bessel_chirped}. For comparison, the Bessel functions that are asymptotic to the solutions as $\rho \to 0$ have been added to both figures. These are given by Eq. \eqref{Delta_eq_0}. When $b>0$ the latter equation suggests a modified Bessel function. The complete evolution of the solution with $b$ is tracked in the map of Fig. \ref{Figure_3}(c). As $b$ assumes lower negative values, the oscillations get compressed because the local transverse wave number increases (recall the discussion following Eq. \eqref{local_wavenumber}). As $b$ assumes larger positive values, the maximum shifts further away from the axis because the caustic radius increases. Exactly as happens with the Airy wave packet, the wave amplitude decays super-exponentially toward the dark side of the caustic and thus the interior of the caustic cylinder contains low optical intentisy.

We can now use the chirped Bessel waves that we have determined numerically as an input to an optical antiguide and investigate their propagation dynamics. At this point it should be realized that, similar to standard Bessel beams in homogeneous media, the solutions of Eq. \eqref{bocher_equation} carry infinite power and thus cannot be realized in their ideal mathematical form. To see this, consider the integral of $U^2(\rho) n(\rho) \rho d\rho$ which is proportional to the power carried by the wave. Then, as discussed in the asymptotics of Section \ref{Wave analysis}, for increasing $\rho$, $U(\rho) \sim \rho^ {-(\nu+2)/4}$ which makes this integral diverge. Hence one needs to consider realistic, finite-energy counterparts which are obtained by a smooth or abrupt apodization of the ideal solutions.

\begin{figure}[t]
\centering \includegraphics[width=1.0\textwidth]{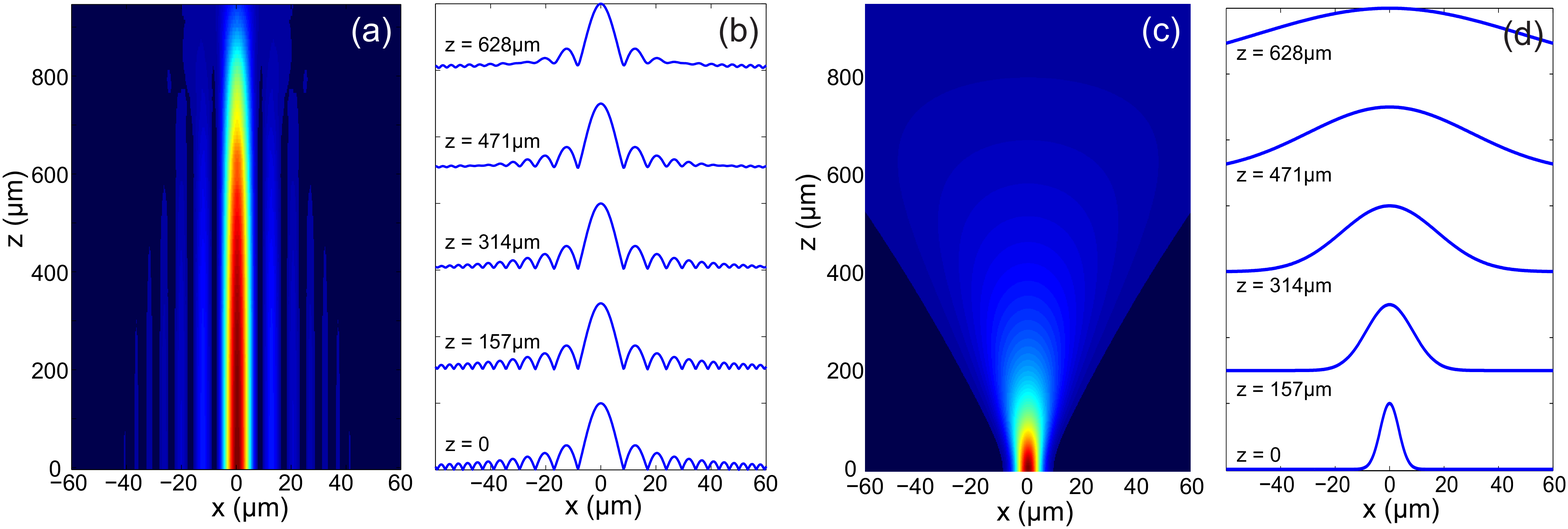}
\caption{(a) Propagation dynamics in terms of the intensity $|U(x,0,z)|^2$ on the $y=0$ plane for a chirped Bessel wave with $m=0$ and $b = -0.001$ (radial profile shown in Fig. \ref{Figure_3}(a)), apodized with the Gaussian $\exp[-(\rho / 125)^2]$ (all distances in $\mu m$), in an antiguide with the parameters of Fig. \ref{Figure_1}. (b) Amplitude profile $|U(x,0,z)|$ of the wave of (a) at different $z$ levels. (c-d) Corresponding results for the Gaussian input wave $U(\rho,z=0) = \exp[-(\rho / 4)^2]$. For visualization purposes, (c) shows the amplitude $|U|$ instead of the intensity $|U|^2$.}
\label{Figure_4}
\end{figure}

Figure \ref{Figure_4}(a) depicts the propagation of a chirped Bessel wave with zero angular momentum $(m=0)$ and a given $k_z$ in a quadratic antiguide. The input condition is apodized with a Gaussian function. The simulation is a split-step Fourier numerical integration of Eq. \eqref{paraxial_equation}. \cite{Agrawal_Book} The result clearly verifies the non-diffracting character of the wave, which is manifested in its apodized version as a remarkable resistance to diffraction. This becomes evident in more detail in Fig. \ref{Figure_4}(b) which shows the radial distribution of intensity at various distances along the antiguide. Notice how the profile, and particularly the central lobe, persists with increasing $z$ despite the defocusing action of the unstable refractive index profile. As visualized with ray optics in Section \ref{Ray analysis}, the outer intensity rings constantly ``feed'' the central lobe with energy thus compensating for the spreading effect of the antiguide. Moreover, the interference of these contributions is such that its shape remains unchanged over an appreciable propagation distance which approaches 1 $mm$ for the given apodization function. To appreciate the uniqueness of this effect even more, in Figs. \ref{Figure_4}(c) and (d) we consider the propagation of a Gaussian input wave with the same full width at half maximum (FWHM). This is similar to launching just the central lobe of the chirped Bessel wave. In the absence of some special structure in its wavefront, such a wave has no chances to survive the defocusing action of the antiguide and thus spreads quickly. These results already suggest that chirped Bessel waves can outperform standard waves (like Gaussian or the simple Laguerre-Gaussian modes used in \cite{Marrucci2013}) in applications where light has to be guided in a diffraction-resisting manner in unstable optical potentials.

\begin{figure}[t]
\centering \includegraphics[width=1.0\textwidth]{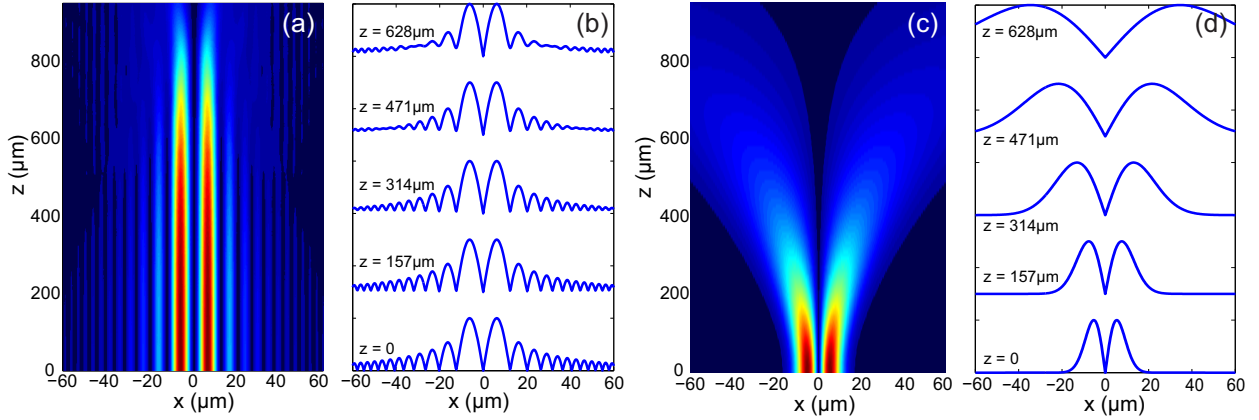}
\caption{(a) Propagation dynamics in terms of the intensity $|U(x,0,z)|^2$ on the $y=0$ plane for a vortex chirped Bessel wave with $m=1$ and $b = -0.001$, apodized with the Gaussian $\exp[-(\rho / 125)^2]$ (all distances in $\mu m$), in an antiguide with the parameters of Fig. \ref{Figure_1}. (b) Amplitude profile $|U(x,0,z)|$ of the wave of (a) at different $z$ levels. (c-d) Corresponding results for the Laguerre-Gaussian input wave $u(\rho,\varphi, z=0) = \rho \exp[i \varphi - (\rho / 7.5)^2]$. For visualization purposes, (c) shows the amplitude $|U|$ instead of the intensity $|U|^2$.}
\label{Figure_5}
\end{figure}

Figure \ref{Figure_5} shows another example, this time for a vortex wave with $m=1$. Subfigures (a) and (b) demonstrate the resistance of this wave to diffraction in the same antiguide with Fig. \ref{Figure_1}. By contrast, in (c) and (d) we observe the strong spreading of a wave with input amplitude of the form $\rho \exp(i\varphi -\rho^2 / w^2)$, namely the focal-plane expression of the lowest Laguerre-Gauss mode with the same order of vorticity. Again, the structure of the chirped intensity rings is what keeps the chirped Bessel wave undistorted along the antiguide, in contrast to the single ring of the Lagueree-Gauss mode which spreads out quickly.

\begin{figure}[t]
\centering \includegraphics[width=1.0\textwidth]{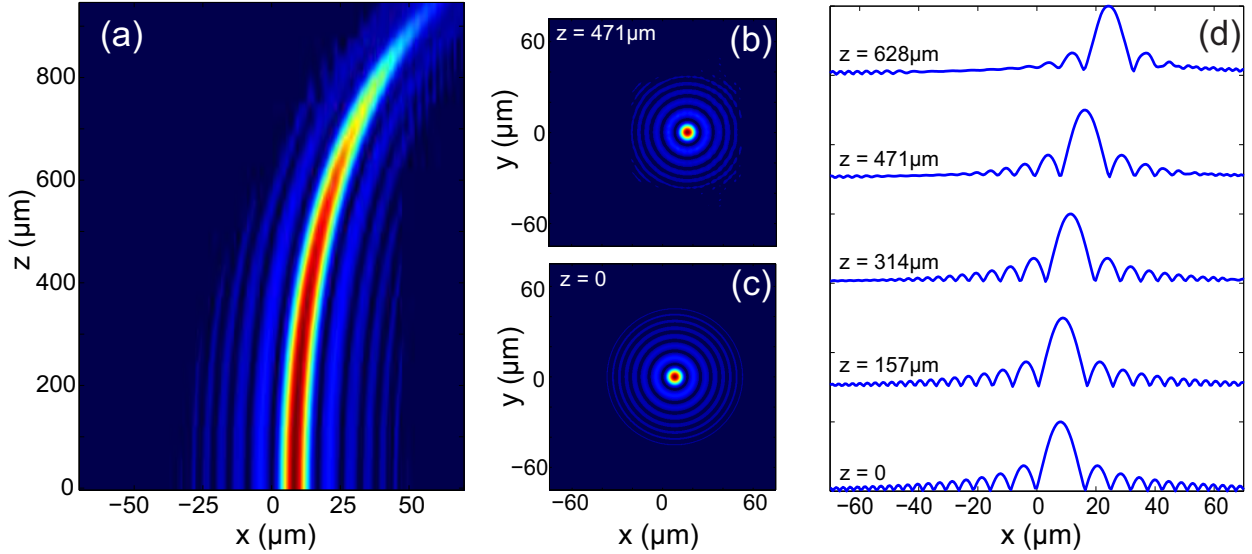}
\caption{(a) Propagation dynamics in terms of the intensity $|U(x,0,z)|^2$ on the $y=0$ plane for the wave of Fig. \ref{Figure_4} launched with a 8 $\mu m$ offset from the antiguide's axis in the $x$ direction. (b-c) Transverse intensity profiles at the shown values of $z$. (d) Amplitude profile $|U(x,0,z)|$ of the wave of (a) at different $z$ levels.}
\label{Figure_6}
\end{figure}

Let us now test our waves under some abnormal conditions. The first such condition is eccentric launching and is examined in Fig. \ref{Figure_6}, where the chirped Bessel wave of Fig. \ref{Figure_4} is launched with its axis displaced by few $\mu m$ from the axis of the antigide. The result is quite remarkable: The wave is attracted toward the high-index region along a curved trajectory in the direction of the displacement ($x$ in this example) while retaining its transverse profile. In the amplitude profiles shown in (d) one can indeed verify the impressive robustness of the main lobe during propagation.

The second adverse condition is launching the wave in a curved antiguide. To emulate the bending effect we solve Eq. \eqref{paraxial_equation} with a $z$-dependent potential of the form $V_b(x,y,z) = V \left( \sqrt{(x-f(z))^2 + y^2} \right)$ where $V$ is the potential defined below Eq. \eqref{paraxial_equation} and $f(z)$ is the function describing the curve of the antiguide, the simplest choice being a parabola, i.e. $f(z) = \gamma z^2$. Chirped Bessel waves of zero (a-b) and nonzero (c-d) angular momentum are examined. In both cases the wave starts propagating along a straight trajectory demonstrating its non-diffracting property. After some distance (around 400 $\mu m$ in both examples) and due to the bending of the antiguide, the wave finds a large part of its wavefront displaced with respect to the antiguide's axis. Thus, as in the case of eccentric launching, its trajectory bends toward the region of increasing index (to the left in these examples). Still, despite this deformation of trajectory the amplitude profile remains fairly robust. Notice the difference of these results with the similar numerical experiment reported in \cite{Marrucci2013} using a simple Laguerre-Gauss mode with vorticity $m=1$. In the latter case and due to the absence of some special structure in its wavefront, the wave gets severely distorted.

\begin{figure}[t]
\centering \includegraphics[width=1.0\textwidth]{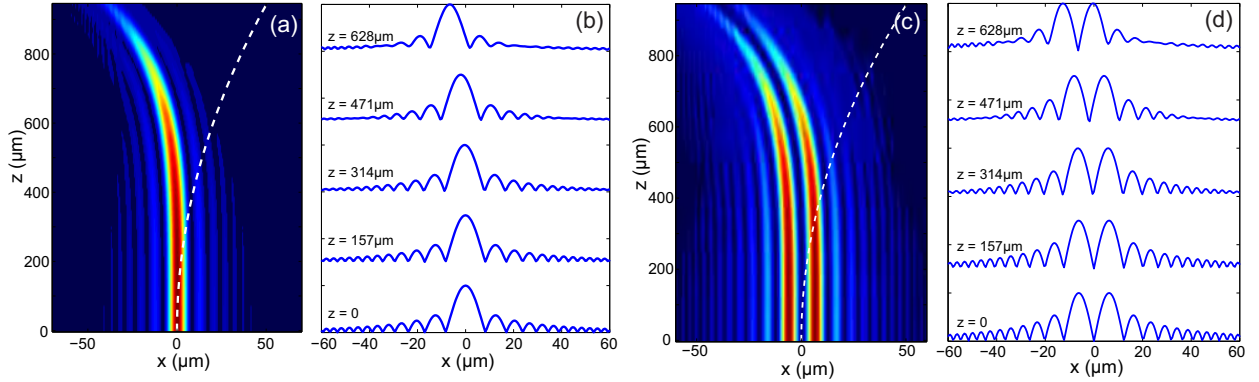}
\caption{(a) Propagation dynamics in terms of the intensity $|U(x,0,z)|^2$ on the $y=0$ plane for the wave of Fig. \ref{Figure_4} launched in an antiguide whose axis is curved along to the parabola $f(z) = (z/133)^2$ (white dashed line, all distances in $\mu m$). (b) Amplitude profile $|U(x,0,z)|$ of the wave of (a) at different $z$ levels. (c-d) Corresponding information for the vortex wave of Fig. \ref{Figure_5} launched in the same curved antiguide.}
\label{Figure_7}
\end{figure}

\begin{figure}[t]
\centering \includegraphics[width=0.8\textwidth]{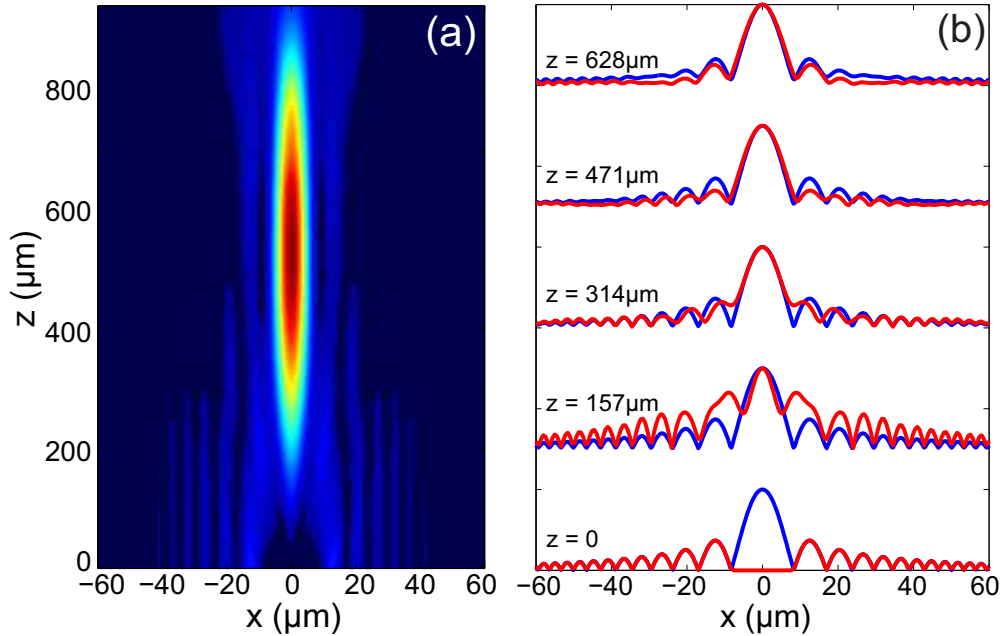}
\caption{(a) Propagation dynamics in terms of the intensity $|U(x,0,z)|^2$ on the $y=0$ plane for the chirped Bessel wave of Fig. \ref{Figure_4} with its central lobed blocked. (b) Amplitude profile $|U(x,0,z)|$ at different $z$ levels for the wave of (a) (red lines) and the wave of Fig. \ref{Figure_4}(a) (blue lines).}
\label{Figure_8}
\end{figure}

The third and last adverse scenario aims to demonstrate self-healing which we have already mentioned as a property of the chirped Bessel waves while discussing their ray optics. An example is shown in Fig. \ref{Figure_8} where the chirped Bessel wave of Fig. \ref{Figure_4} is launched in the same quadratic antiguide with its main lobe removed. Despite this severe distortion, the wave manages to reconstruct the main lobe and part of the profile after some distance.

\begin{figure}[t]
\centering \includegraphics[width=1\textwidth]{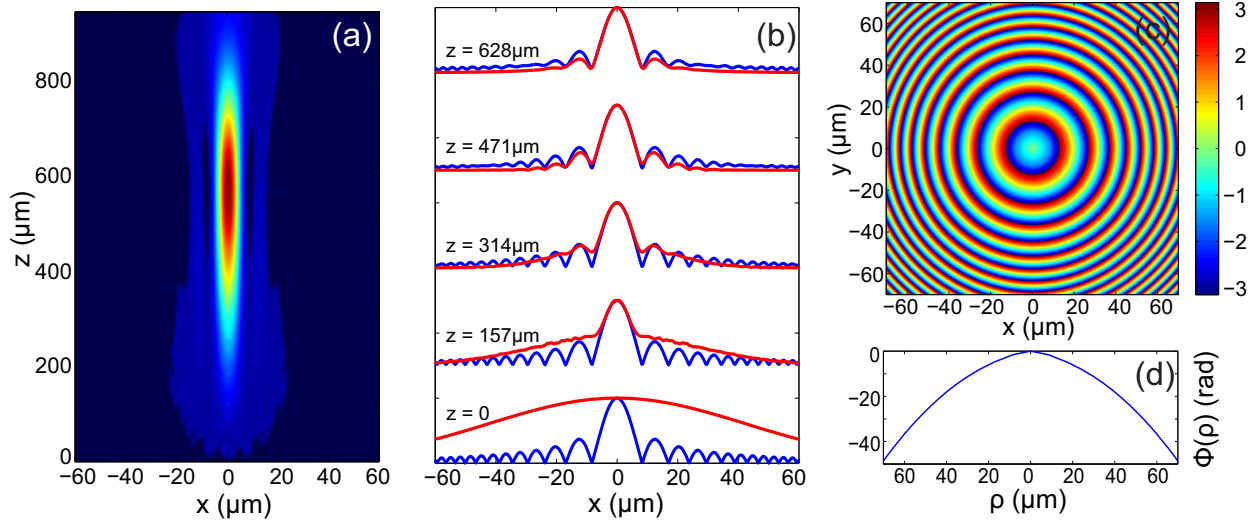}
\caption{(a) Propagation dynamics in terms of the intensity $|U(x,0,z)|^2$ on the $y=0$ plane for the input wave $\exp[i \Phi(\rho) - (\rho / 60)^2]$ launched in the antiguide of Fig. \ref{Figure_1} (all distances in $\mu m$). (b) Amplitude profile $|U(x,0,z)|$ at different $z$ levels for the wave of (a) (red lines) and the wave of Fig. \ref{Figure_4}(a) (blue lines). (c) The modulo-2$\pi$ phase function $\Phi(x,y)$ that modulates the Gaussian wave. (d) The same function (in radians) versus the radius.}
\label{Figure_9}
\end{figure}

We close with a more practical approach to the above effects. In practice, it may be difficult to realize an exact wave function experimentally. This might be true not only for our chirped Bessel waves but for any kind of structured lightwave. It is thus always useful to come up with simpler input waves able to approximate the same effect which in our case is the diffraction-resisting propagation in an optical antiguide. From a closer look at the ray patterns in Figs. \ref{Figure_2}, one sees that, at any given $z$, there are two rays passing from a point on the $\rho$ axis, an inward and an outward one, with equal but opposite slopes with respect to the z axis. Of the two, it is the inward ray that contributes to the field near the axis at larger values of $z$. This implies that the profile of the wave near the axis will not be affected significantly if we discard the outward rays from the input condition at $z=0$. Such a condition can be easily implemented in the laboratory with a circularly symmetric phase mask (or a spatial light modulator) that has the appropriate phase function $\Phi (\rho)$ to create the inward rays with the correct slope. $\Phi (\rho)$ is obviously the phase of the oscillations of the solution of Eq. \eqref{bocher_equation} (assuming that it oscillates proportionally to $\cos(\Phi (\rho))$ and the transmission function of the phase mask should be $\exp[-i \Phi (\rho)]$, with the negative sign giving rise to inward rays. An relevant example is given in Fig. \ref{Figure_9}, where we are trying to approximate the dynamics of Fig. \ref{Figure_4}(a) with a Gaussian input wave modulated by the phase mask shown in (c-d). For a better comparison, (b) shows the amplitude profile at several values of the propagation distance superposed on the corresponding profiles of Fig. \ref{Figure_4}(b). The goal is obviously achieved as the phase-modulated Gaussian wave reproduces almost excellently the main lobe and the first intensity ring of the chirped Bessel wave, within a certain range of $z$.

\section{Conclusion}
We have introduced and studied chirped Bessel waves capable of resisting diffraction in optical antiguides with a power-law varying index of refraction. We examined the ray structure of these waves with a detailed (almost tutorial) spirit and investigated their propagation dynamics with full-wave split-step Fourier simulations. Conditions of eccentric illumination, antiguide bending and partial obstruction of the input wavefront were also examined and demonstrated the robustness of their profile and the ability for self-healing. A practical implementation of these waves using simple phase-modulated input waves was finally proposed. From a more general viewpoint, this work adds to the family of structured lightwaves members that resist (or completely overcome, in their ideal mathematical version) diffraction in media that are inherently more diffractive than free space, such as unstable optical ducts. And we have done so by extending the family of Bessel beams with their radially chirped counterparts. The former are non-diffracting in free space while the latter in optical antiguides.


\end{document}